\documentclass[doublecol]{epl2}
\usepackage{graphicx}
\usepackage{amsmath,amssymb,revsymb,epsfig}
\usepackage{color}
\usepackage{multirow}
\usepackage{float}
\usepackage{bm}
\usepackage{bbm}
\usepackage{epstopdf}

\title{Can a photonic thermalization gap arise in disordered non-Hermitian Hamiltonian systems?}

\author{Guang-Lei Wang\inst{1}, Hong-Ya Xu\inst{1}, Ying-Cheng Lai\inst{1,2}}

\institute{
\inst{1}School of Electrical, Computer and Energy Engineering, Arizona State University, Tempe, Arizona 85287, USA\\
\inst{3}Department of Physics, Arizona State University, Tempe, Arizona 85287, USA
}

\date{\today}

\abstract{
The phenomenon of chiral-symmetry protected thermalization gap in Hermitian 
photonic systems is counterintuitive as it implies that the photon coherence 
can be continuously improved by disorders towards an asymptotic limit. We show 
that the phenomenon disappears in time-independent, non-Hermitian photonic 
systems even when the chiral symmetry is well preserved. In fact, the degree 
of thermalization generally increases with the disorder strength, in agreement
with intuition. As non-Hermitian characteristics (e.g., weak gain and loss)
can be expected in realistic physical situations, the phenomenon of 
thermalization gap may be observed but only in well controlled experiments 
with high quality materials.}


\begin{document}
\maketitle

{\bf\it Introduction}. 
Recent years have witnessed a rapid growth of interest in non-Hermitian
systems~\cite{CW:2015,KYZ:2016}. In open Hamiltonian systems with symmetry
breaking so that the Hermitian properties can no longer be maintained,
surprising physical phenomena can arise~\cite{CEPH:2013,RMBNOCP:2013,
LRM:2014,LC:2014,MPS:2015,SGWC:2015}. For example, in condensed matter
physics, the principle of bulk-boundary correspondence
stipulates that the bulk of the lattice can be characterized by a
topological invariant whose value determines the existence of possible
gapless edge states. However, a recent work~\cite{L:2016} demonstrated that
violating the Hermitian properties can induce a fractional topological
invariant number and lead to a change in the stability of the edge states,
thereby defying the bulk-boundary correspondence. Another example occurs
in wave propagation/transport, where conventional wisdom is that
phenomena such as diffusion and localization are caused by random disorders.
However, it was demonstrated theoretically and
experimentally~\cite{EHWSDCNS:2013,GWJNLBEMS:2014} that a sudden transition
from ballistic to diffusive motions can occur even in ordered, 
time-independent, non-Hermitian systems maintaining the parity-time reversal
($\mathcal{PT}$) symmetry. Also, a remarkable phenomenon in
non-Hermitian systems is the appearance of exception points in the eigenvalue
spectrum about which crossing/anti-crossing transitions can occur and the 
corresponding emergence of a new topological Berry phase associated with
paths encircling the exceptional points~\cite{GEBLFBKSHY:2015,LGCSTR:2012}.
In physical contexts such as resonances and discrete-continuous energy
transitions, the eigenvalues of the Hamiltonian are necessarily complex
to account for, e.g., the finite lifetime of the state. Similarly, in 
non-Hermitian systems, the eigenvalues are complex. Experimentally, 
non-Hermitian physics can be implemented and observed in waveguides, photonic 
crystals, dielectric micro-resonators, and even in biological neutral 
networks~\cite{AHN:2016}.

In one-dimensional Hermitian systems, the emergence of a chiral symmetry 
protected photonic thermalization gap has been theoretically 
predicted~\cite{KAS:2015,KAS2:2015,KSACS:2016} and observed
experimentally~\cite{KKSACS:2016,KSACS:2016,KAS:2017}. A system is said to 
possess a chiral symmetry if (a) there are eigenvalues (labeled by integers 
$m$ and $-m$) that appear in pairs whose real and imaginary parts have 
opposite signs, and (b) the associated eigenstates satisfy the relation: 
$\phi^m_n = (-1)^{n}\phi^{-m}_n$, where $n$ is the space coordinate. Of 
interest is how random disorders affect the statistical properties of photon 
thermal fluctuations. To measure the statistical properties of these 
fluctuations, a second-order coherence function, i.e., the normalized 
intensity correlation $g^{(2)}$ [Eq.~\ref{eq:g2} below], is commonly 
used~\cite{KAS:2015,KAS2:2015,KSACS:2016,Szameit:2015}, which is capable of 
revealing statistical information about the source. The values of $g^{(2)}$ 
are one for fully coherent light (e.g., laser) and two for random or chaotic
light (e.g., black body radiation). In the absence of any disorder, there is 
a high degree of coherence among the photons. In this case, $g^{(2)}$ assumes 
the unity value corresponding to coherent photon states. When there are 
arbitrarily weak disorders, i.e., when the strength of disorder is turned on 
from zero, there is an abrupt transition in $g^{(2)}$ from unity to a value 
corresponding to incoherence at which the photon statistics can be 
approximately described by the modified Bose-Einstein distribution. As the 
disorder strength is further increased, the value of $g^{(2)}$ decreases but 
can never reach the unity value again in systems with a chiral symmetry. In 
fact, the lower bound of the measure has the value of two at which the 
Bose-Einstein distribution holds. A ``thermalization gap'' thus arises in 
the plot of the measure versus the disorder 
strength~\cite{KAS:2015,KAS2:2015,KSACS:2016}.

The emergence of a photon thermalization gap~\cite{KAS:2015,KAS2:2015,
KKSACS:2016,KSACS:2016,KAS:2017} is striking and counterintuitive, as it 
implies that, beyond the regime of weak disorder, the degree of photon 
coherence in the underlying system can be continuously improved by increasing 
the disorder strength. This result was obtained for purely Hermitian 
systems with a chiral symmetry, e.g., a photonic system of a set of ideal 
parallel coupled waveguides with the values of the refractive index being 
purely real, where there is no emission and/or absorption. In physical
reality the refractive index of any material can be expected to be generically 
complex with an inevitable nonzero imaginary part. This consideration
motivated us to investigate thermalization in non-Hermitian photonic systems.

To choose a prototype system setup, we note that, in principle, we can 
impose certain symmetries on the non-Hermitian system. However, special 
arrangement of the configuration may be required. For example, if we demand 
that the system be time independent and possess a $\mathcal{PT}$ symmetry, 
the real and imaginary parts of the refractive index must be an even and odd 
function in space, respectively~\cite{MECM:2008,MAKKC:2012}. Although 
$\mathcal{PT}$ symmetry without such a special arrangement can be realized, 
time modulation or spatial engineering of the material along the propagation 
direction is required~\cite{LHZQXKL:2013}. We are thus led to consider the 
more general setting where we assume that the non-Hermitian system is time 
independent and does not possess a $\mathcal{PT}$ symmetry. Specifically, 
we study the photon thermalization statistics in photonic systems of 
waveguide array in which the refractive index of each individual element 
can be complex with either sign (i.e., gain or loss). To be more general, 
we consider disorders that can either retain or break the chiral symmetry. 
We develop a framework of coupled mode equations for non-Hermitian waveguide 
systems with varying separations and refractive indices, which enables us
to address the question of whether a chiral-symmetry protected thermalization 
gap can arise in non-Hermitian photonic systems. The main finding is that
the degree of photon coherence continues to deteriorate as the disorder 
strength is increased, a result in sharp contrast to that in
Refs.~\cite{KAS:2015,KAS2:2015,KSACS:2016}. Our work thus establishes
the necessary conditions under which a thermalization gap can be expected in 
a photonic system. In particular, unless in a well controlled experiment 
with nearly perfect materials in the intermediate time (propagation distance) 
regime, the counterintuitive phenomenon that disorder can continue to 
improve the coherence of photon fluctuations cannot occur. In fact, in 
non-Hermitian systems no thermalization gap can be expected.

{\bf\it Non-Hermitian photonic systems}.
Hermitian symmetry requires $H=H^\dagger$ and the system is chiral
symmetric if $CHC = -H$, with $C=diag(1,1,\dots,1,-1,\dots,-1)$ being the
chiral symmetry operator. In general, the Hamiltonian matrix of a photonic
waveguide system can be expressed in a $2\times 2$ block form:
\begin{equation*}
    H \rightarrow
    \left[
        \begin{array}{cc}
        H' & H_S \\ H_S^T & H''
        \end{array}
    \right],
\end{equation*}
where $H', H''$ and $H_S$ are matrices in the corresponding blocks. For a
system with chiral symmetry, the diagonal blocks are zero: $H' = H'' = 0$.
Further, if the system is Hermitian, the left-bottom block can be expressed 
as $H_S^\dagger$. In this case, the Hamiltonian is real, which is the 
class of systems in which the phenomenon of a thermalization 
gap~\cite{KAS:2015,KAS2:2015,KKSACS:2016,KSACS:2016,KAS:2017} was uncovered. 
For systems considered in our work, we have $H_S^T \neq H_S^\dagger$, and 
the Hamiltonian is in general complex.

\begin{figure}
\centering
\includegraphics[width=\linewidth]{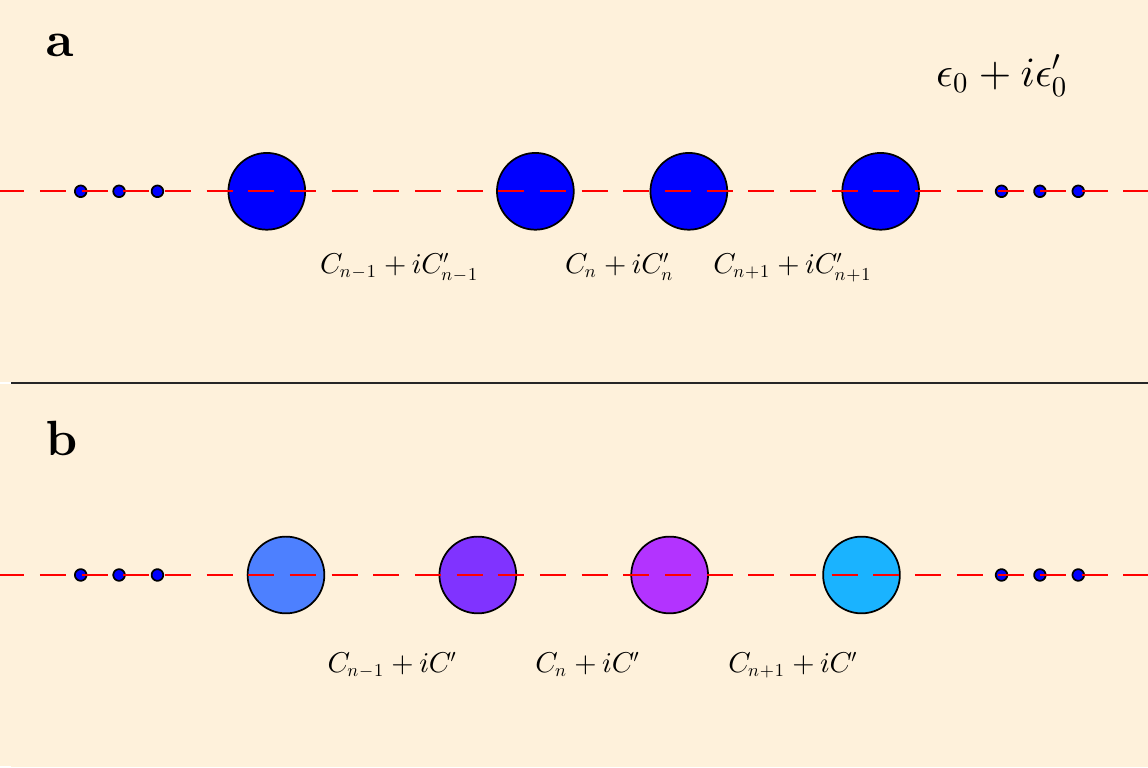}
\caption{ {\bf Two types of disorders in a non-Hermitian
photonic system of a waveguide array}. Both the waveguides and the
surrounding medium have a complex reflection index. (a) Spacing disorder
in which the distances between the nearest waveguides are randomly modified.
The disorders will result in randomness in both the real and imaginary
parts of the coupling constant. (b) Relative permittivity disorder in
which the complex refractive index of each waveguide is randomly modified
while the distances between the nearest waveguides are kept constant. As a
result, there will be random perturbations upon both the diagonal and
off-diagonal elements of the Hamiltonian.}
\label{fig:array}
\end{figure}

As a concrete setting to investigate the photon statistics in non-Hermitian
systems, we consider an array of one-dimensional, single-mode optical
waveguides arranged in the $x$ direction with gain and loss, as shown in 
Fig.~\ref{fig:array}. Each waveguide has complex relative permittivity 
$\epsilon_n+i\epsilon_n'$, and the waveguides are placed in a surrounding 
medium with complex permittivity $\epsilon_0+i\epsilon_0'$. In the idealized 
situation where the waveguides are identical and equally spaced, the system 
can be described by a set of complex, coupled mode 
equations~\cite{GWJNLBEMS:2014}. In a realistic situation, random factors 
are present, which introduce disorders that can be modeled as proper 
perturbations to the coupled mode equations.

We start from the Helmholtz equation governing wave propagation in the system:
\begin{equation} \label{eq:Helmholtz}
[\nabla^2+k_0^2\tilde{\epsilon}(x)]\Psi(x,z)=0,
\end{equation}
where $z$ is the propagation direction, $\Psi(x,z)$ is the electric field 
amplitude, $k_0$ is the free space wave vector, and $\tilde{\epsilon}(x)$ is 
the profile of the relative electric permittivity of the system, which can be 
expressed as
\begin{equation}
\tilde{\epsilon}(x)=\epsilon_0+i\epsilon_0'+\sum_{n=1}^N
[(\epsilon_n-\epsilon_0)+i(\epsilon_n'-\epsilon_0')]\zeta_n(x),
\end{equation}
where $2w$ is the width of the each waveguide and
$\zeta_n(x)=\Theta(x-x_n+w)-\Theta(x-x_n-w)$ with $\Theta(x)$ being
the Heaviside step function. The effective coupled mode equations can be
obtained under the approximation that the eigenmode of the individual
waveguide is well confined so that the full width half maximum (FWHM)
$\ell$ is much smaller than $d$, the spacing between two adjacent
waveguides. The equations are~\cite{GWJNLBEMS:2014}
\begin{equation}
-i\frac{d\phi_n}{dz}=i\kappa\phi_n+(C+iC')(\phi_{n+1}+\phi_{n-1}),
\end{equation}
where
\begin{eqnarray} \label{eq:factor}
\kappa&&=\frac{k_0^2}{2\beta}[\epsilon_0'
+(\epsilon'-\epsilon_0')\tanh(\frac{w}{\ell})], \nonumber \\
C&&=\frac{(\epsilon-\epsilon_0)k_0^2}{2\beta}
\frac{w}{\ell}\exp{(-\frac{d}{\ell})}, \\
C'&&=\frac{(\epsilon_0'-\epsilon')k_0^2}{\beta}
\frac{d}{\ell}\exp{(-\frac{d}{\ell})}. \nonumber
\end{eqnarray}
The quantity $\phi_n$ in Eq.~(\ref{eq:factor}) is the field amplitude in the
$n$th waveguide and $k_0\sqrt{\epsilon_0}<\beta<k_0\sqrt{\epsilon}$ is the
propagation constant along the $z$ direction.

To incorporate disorder-immune chiral symmetry in the non-Hermitian waveguide
system, we consider two types of disorders. For the first type, the distance
between a pair of adjacent waveguides is randomly modified, as shown in
Fig.~\ref{fig:array}(a). Mathematically, insofar as the approximation
$\ell\ll |x_n-x_{n-1}|$ holds, the general form of the coupled mode equations
remains the same, except that the coupling constant now has a spatial
dependence:
\begin{eqnarray} \label{eq:factor2}
\nonumber
	C\rightarrow C_{n,n\pm1}&=&\frac{(\epsilon-\epsilon_0)k_0^2}{2\beta}
\frac{w}{\ell}\exp{(-\frac{|x_n-x_{n\pm1}|}{\ell})}, \\
	C'\rightarrow C'_{n,n\pm1}&=&\frac{(\epsilon_0'-\epsilon')k_0^2}{\beta}
\frac{|x_n-x_{n\pm1}|}{\ell}\cdot \\ 
	& & \exp(-\frac{|x_n-x_{n\pm1}|}{\ell}), \nonumber
\end{eqnarray}
where $|x_n-x_{n\pm1}|=d(1+\xi_1)$ and $\xi_1$ is a random variable uniformly
distributed in $[-\Delta C, \Delta C]$. We consider the weak disorder regime
and retain the first order corrections. In general, random modulation in the 
waveguide spacing leads to disorders in both the real and imaginary parts of 
the coupling constant. In a matrix representation of the coupled mode 
equation, this type of disorders will be present in the off-diagonal elements 
only, hence the term ``off-diagonal disorders.'' In spite of the presence of 
waveguide spacing disorders, the system is still chiral-symmetric.

While the off-diagonal disorders described above preserve the chiral
symmetry of the non-Hermitian system, the second type of disorders we
study break the chiral symmetry. Conventionally, this can be done
by introducing disorders into the diagonal elements of the coupling
matrix. However, as can be seen from Eq.~(\ref{eq:factor2}), it is not
feasible to introduce diagonal disorders while keeping the off-diagonal
elements unperturbed. Our strategy is then to randomly modify the real
part of the relative permittivity of the individual waveguides while
keeping the imaginary part identical. This way both diagonal and
off-diagonal disorders are present to break the chiral symmetry, as
shown in Fig.~\ref{fig:array}(b). Mathematically, including this kind of
disorders entails altering the quantities $\kappa$ and $C$ in the
following manner:
\begin{eqnarray} \label{eq:nHfactor}
i\kappa\rightarrow&& i\kappa+\frac{(\epsilon_n-\bar{\epsilon})k_0^2}
{2\beta}\tanh{(\frac{w}{\ell})}, \\
C\rightarrow&& C_{n,n\pm1}=\frac{(\epsilon_n+\epsilon_{n\pm1}-\bar{\epsilon}
-\epsilon_0)k_0^2}{2\beta}\frac{w}{\ell}\exp{(-\frac{d}{\ell})} \nonumber,
\end{eqnarray}
where $\epsilon_n$ is the real relative permittivity of the $n$th waveguide
and $\bar{\epsilon}$ is the corresponding average value of all the waveguides.
We set $\epsilon_n=\epsilon(1+\xi_2)$, where $\xi_2$ is a random variable
uniformly distributed in $[-\Delta\beta,\Delta\beta]$.

To characterize the photon statistics in the non-Hermitian system, we
use the following three quantities~\cite{KAS:2015}: the normalized
intensity correlation $g^{(2)}(z)$, the intensity probability distribution
$P(\bar{\mu})$, and the photon-number distribution $P(n_{ph})$. In
particular, $g^{(2)}(z)$ characterizes the degree of randomness of 
light in the probability space of disorder realizations, which is defined
as~\cite{KAS:2015}
\begin{equation} \label{eq:g2}
g_x^{(2)}(z)=\frac{\langle I^2_x(z)\rangle}{\langle I_x(z)\rangle^2},
\end{equation}
where $I_x(z)=|\phi_x(z)|^2$ is the intensity of the electric field at the
$x$th waveguide and $\langle\cdot\rangle$ denotes ensemble averaging. For
coherent light, we have $g^{(2)}=1$. For thermalized (random or chaotic)
light, we have $g^{(2)}=2$, where the intensity correlation can be
evaluated at the middle of the waveguide array~\cite{KAS:2015,KAS2:2015}.
One may also consider the normalized intensity correlation between a pair
of waveguides, from which non-Gaussian statistics corresponding to photon
antibunching have recently been found~\cite{KMKPSACS:2016}.

The intensity probability distribution $P(\bar{\mu})$ and the full
photon-number distribution $P(n_{ph})$ can exhibit different scaling
behaviors for different degrees of photon thermalization, e.g.,
coherent, thermalized/chaotic and super-chaotic. In our study, we
set the initial excitation at the center of the waveguide array and
generate a large number of realizations of the input intensity, which 
obey the coherent Poisson distribution. After the excitation propagates
in the waveguides for certain distance (which is effectively time), we
examine the intensity at the central waveguide, which is proportional to
the photon numbers. The photon-number distribution can be obtained through
the ensemble statistics after considering the effect of measurement.

For a Hermitian system, the conventional solution approach is to expand the
initial wavefunction in terms of the eigenfunctions and then to calculate the
time evolution of each eigenfunction. However, this approach is not suitable
for non-Hermitian systems, as the eigenfunctions of a non-Hermitian matrix
are generally not orthogonal to each other. We thus resort to direct 
numerical solutions of the coupled mode equations. For convenience, we use 
the same parameters as in the experimental work~\cite{GWJNLBEMS:2014}: 
$C_0=0.1\mbox{mm}^{-1}$ to rescale all the parameters, and set
\begin{equation}
\alpha=2\frac{(\epsilon_0'-\epsilon')}{(\epsilon-\epsilon_0)}\frac{d}{w}=-0.15
\end{equation}
to relate $C'$ to $C_0$ and $\kappa/C_0=0.3$. For adjacent distance disorders,
since the unaffected diagonal elements can be removed by incorporating
the quantity $e^{-\kappa t}$ into all the wave amplitudes, i.e.,
$\phi(z)=E(z)e^{-\kappa t}$. This extra time evolution term contributes
to an overall gain or loss, depending on the sign of $\kappa$, but it is
identical for all disorder realizations, which can be canceled out when
calculating $g_0^{(2)}$. For disorders in the real relative permittivity,
since the imaginary part is constant, the diagonal $i\kappa$ term can still
be removed. Other parameters are set to be $d=17\mu\mbox{m}$, $2w=5\mu\mbox{m}$,
$\epsilon_0/\epsilon=0.2$, and $d/\ell\sim2$.

We choose the initial condition as a single excitation at the central 
waveguide. For a Hermitian system, this choice guarantees that both modes in 
each chiral pair have the same weight during propagation, activating the 
chiral symmetry. However, in a non-Hermitian system, orthogonality and 
completeness are not guaranteed for the eigenvectors. While it may be 
possible to design an input excitation for any specific disorder realization 
so that the symmetric-excitation condition is fulfilled, the gain/loss 
effects would induce imbalances between each pair of modes during 
propagation, defying the symmetric-excitation condition.

\begin{figure}
\centering
\includegraphics[width=\linewidth]{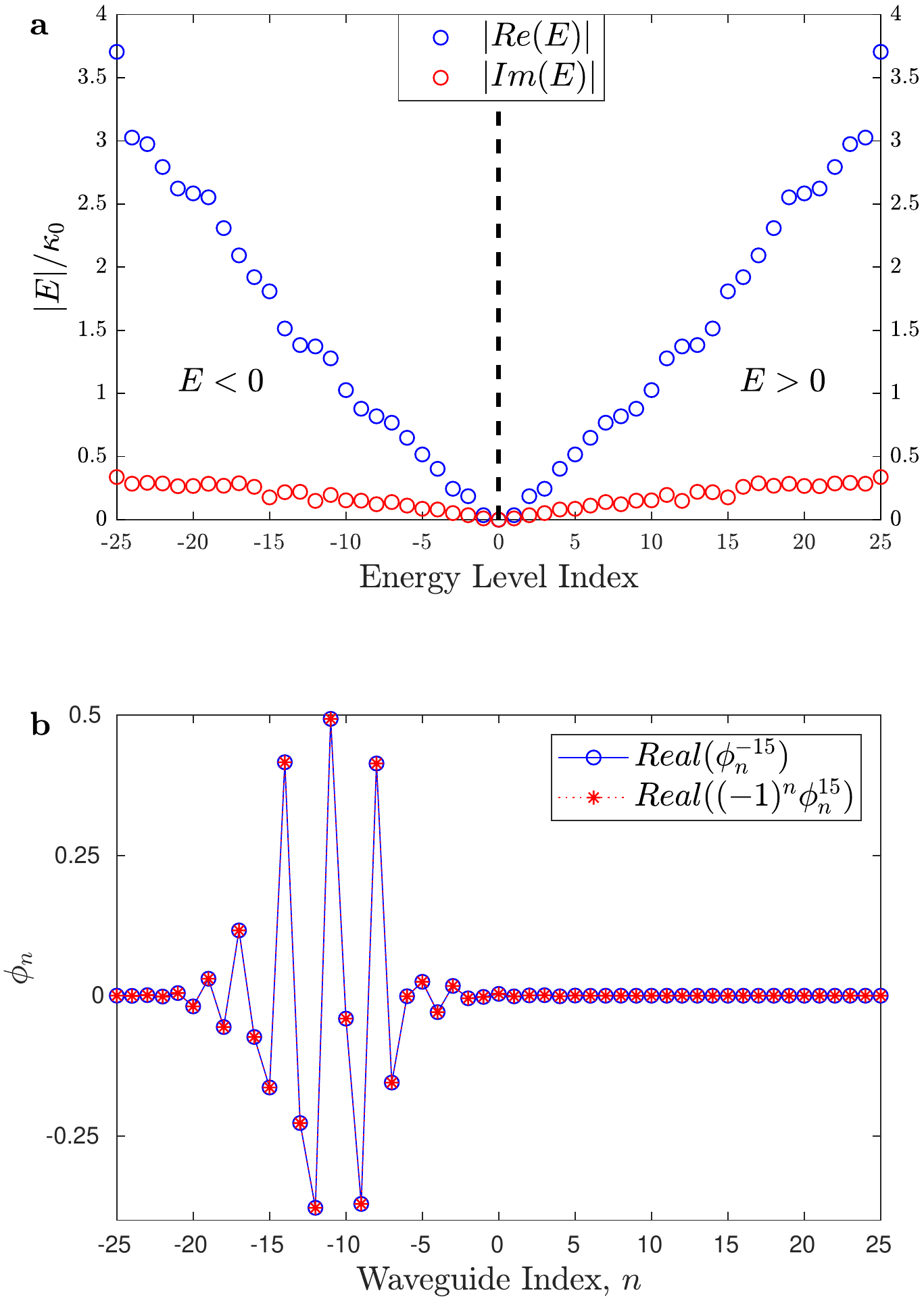}
\caption{ {\bf Properties of non-Hermitian systems with off-diagonal
disorder}. (a) Complex eigenvalue spectrum exhibiting an odd symmetry
for both the real and imaginary parts: $\xi_m=-\xi_{-m}$. It is convenient
to take the absolute values of the eigenvalues. The energies in the left
panel are negative. (b) A representative pair of eigenfunctions with the 
property $\phi^m_n=(-1)^n\phi^{-m}_n$. Shown are the real parts of the 
eigenfunctions with $m=-15$. The imaginary parts share the same symmetry.}
\label{fig:eigen}
\end{figure}

{\bf\it Main results}.
To verify the chiral symmetry in our non-Hermitian system, we examine
the eigenvalue spectrum and the associated eigenfunctions for a typical
realization of the off-diagonal disorder, as shown in Fig.~\ref{fig:eigen}.
Mathematically, the chiral symmetry requirement $CHC=-H$ leads to
$\xi_m=-\xi_{-m}$ and $\phi^m_n=(-1)^n\phi^{-m}_n$,
where $\xi_m$ stands for the real or imaginary part of the $m$th eigenvalue
and $\phi^m_n$ is the corresponding eigenfunction at the $n$th waveguide.
Figure~\ref{fig:eigen} reveals a continuous existence of the chiral symmetry
even in the presence of off-diagonal, complex disorders. An alternative way to
test the chiral symmetry is to calculate the commutation between the
Hamiltonian and the chiral symmetry operator. We note that, if the disorders
occur in the real relative permittivity, the chiral symmetry will be broken.

\begin{figure}
\centering
\includegraphics[width=\linewidth]{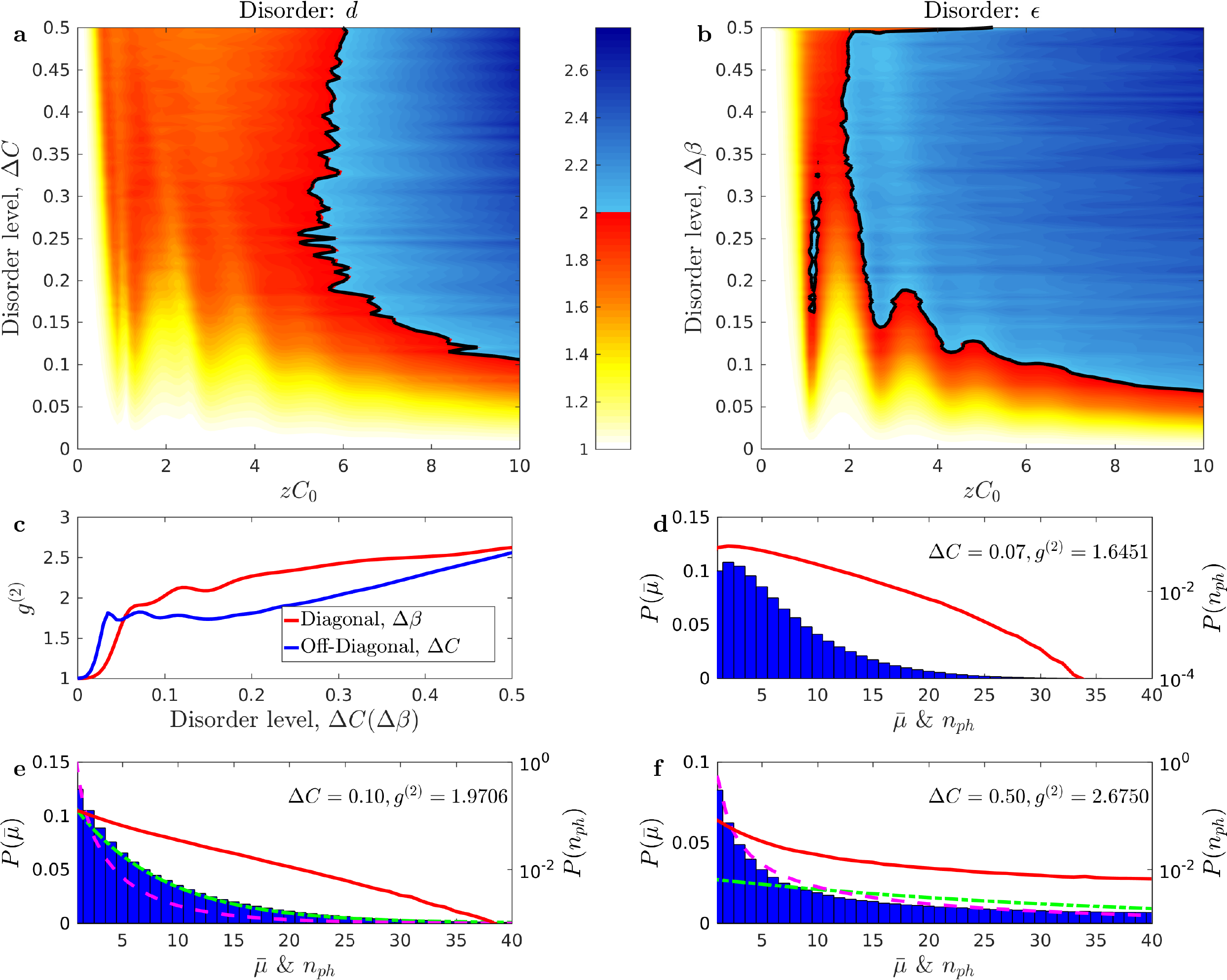}
\caption{ {\bf Behavior of photon thermalization in non-Hermitian systems
subject to random disorders}. (a,b) Normalized intensity correlation
$g_0^{(2)}(z)$ at the center of the waveguide array as a function of the
disorder strength and propagation distance $z$ for a system with adjacent
distance disorder (off-diagonal disorder) and real relative permittivity
disorder, respectively. The number of waveguides in one array is 51.
(c) The correlation measure $g_0^{(2)}(zC_0=10)$ as a function of the
disorder strength for off-diagonal (blue) and permittivity (red)
disorders. (d-f) Intensity probability distribution $P(\bar{\mu})$ and
photon number distribution $P(n_{ph})$ for a coherent input with a
fixed mean photon number $\bar{\mu}=680$ at the input for different values
of the off-diagonal disorder strength. The intensity $I=|\phi|^2$ is
represented by the mean photon number $\bar{\mu}$ while the initial photon
number is set to be $I=1$. The left axis indicates $P(\bar{\mu})$ on a linear
scale (the blue bar) while the right axis is a semi-logarithmic representation
of $P(n_{ph})$ (red). The dashed-dot and the dashed curves 
represent the fitted exponential and Gaussian-square distributions,
respectively, in which the disorder strengths and the corresponding
$g^{(2)}$ values are inserted. Due to the system's being non-Hermitian,
eigenvalues are complex. Numerically it is necessary to set a finite (albeit
relatively large) distance to prevent energy from diverging.}
\label{fig:main}
\end{figure}

Figure~\ref{fig:main} summarizes our main finding that in non-Hermitian
systems, the previously reported chiral symmetry protected photonic
thermalization gap disappears. In particular, Figs.~\ref{fig:main}(a,b)
show the dependence of the normalized intensity correlation $g_0^{(2)}$ on
the propagation distance and the disorder strength for the two types
of disorders. In both cases, a long propagation distance and a high
disorder strength can lead to a relatively high value $g_0^{(2)}$. This
can be understood by noting that the main source of the randomness in the
intensity is the gain/loss that behaves as
$\sim \exp{(\Delta[Im(E_n)]\cdot t)}$.
A higher level of disorder thus leads to a larger value of $\Delta[Im(E_n)]$,
while a longer propagation distance corresponds to a longer time. For the
second type of disorders (i.e., real relative permittivity disorders),
the values of $g_0^{(2)}$ are generally larger as compared to those with
the first type of disorders (i.e., adjacent distance or off-diagonal
disorders), due to the fact that a nonzero value of $\Delta\beta$ induces
both diagonal and off-diagonal disorders. We observe that, in the small time
regime, the main signatures of the Hermitian characteristics of the system as
reported in Ref.~\cite{KAS:2015} are retained, i.e., $g_0^{(2)}$ tends to
decrease with the disorder strength. However, in the long time (or
equivalently, large distance) regime, $g_0^{(2)}$ increases with the
disorder strength. In this regime, the non-Hermitian characteristics
dominate, as the main contributions of the imaginary parts of the
eigenvalues are from the exponential term. As a result, an increase in the
disorder strength will be amplified by the propagation distance, leading
to a continuous increase in the value of $g_0^{(2)}$.

Figure~\ref{fig:main}(c) shows the behavior of $g_0^{(2)}$ versus the
disorder strength for $zC_0=10$. We observe a non-decreasing behavior,
where the value of $g_0^{(2)}$ changes continuously from unity to about
$3$. For longer propagation length, the value of $g_0^{(2)}$ can exceed
$3$, due to the non-Hermitian nature of the system. Say we consider $R$
realizations of the off-diagonal order. Except for the overall gain/loss
factor $\kappa$, the imaginary parts that govern the evolution of the
eigenmodes are random. After a long time, the difference in the intensities
of the eigenmodes grows exponentially, resulting in significant differences
in the intensities of the central waveguide among different realizations.
In such a case, we have
\begin{equation}
g_0^{(2)}(z) = \frac{\langle I_0^2(z)\rangle}
{\langle I_0(z)\rangle^2}
\simeq \frac{I_{0,max}^2(z)/R}{(I_{0,max}(z)/R)^2} =R,
\end{equation}
indicating that $g_0^{(2)}$ can approach the upper bound $R$.

Figures~\ref{fig:main}(d-f) show the full photon number statistics for
the case of off-diagonal disorders for $zC_0=10.0$. From the initial
condition, we generate an ensemble of input intensity, with the distribution
which can approximately be described by the Poisson form:
$P(n)=\bar{\mu}^ne^{-\bar{\mu}}/n!$ with the mean photon
number $\bar{\mu}\approx680$. Due to gain, the effective electric field
$E(z)$ can reach a large value. Since $\phi(z)=E(z)e^{-\kappa t}$,
the real electric field amplitude can still be moderate for an appropriate
value of $\kappa$. Note that the expression of $\kappa$ does not depend
on the nature of the disorders. We can thus adjust the value of $\kappa$
to ensure that the energy of the system has some reasonable value
within a finite distance. Experimentally, the range of the accessible
imaginary relative permittivity can be large, where gain can be realized
e.g., through photon pumping while loss can be introduced by metal.

Figure~\ref{fig:main}(d) shows a case where $g_0^{(2)}=1.6451$ and there is
a transition from the coherent Poisson distribution to the thermalized Gaussian
distribution. About the normal thermal state for which $g_0^{(2)}\approx 2$,
the intensity distribution is exponential: $P(I)=(1/\mu)e^{-I/\mu}$, where
$\mu=\langle n_{ph}\rangle$ is the ensemble averaged photon number and the
photon number statistics follow the Bose-Einstein
distribution~\cite{KAS:2015}. For the superthermal state
$g_0^{(2)}\approx3$, the intensity should follow a Gaussian-square type of
distribution: $P(I)=(1/\sqrt{\pi\mu I})e^{-I/\mu}$,
which is associated with a modified Bose-Einstein photon number
statistics~\cite{KAS:2015}. These behaviors have indeed been observed.
In particular, in Figs.~\ref{fig:main}(e,f),
the left side axis is for the blue bar and is on the linear scale of
$P(\bar{\mu})$, and the right axis is for the red curve $P(n_{ph})$ on
a semi-logarithmic scale, where the mean photon number $\bar{\mu}$ is used
to mark the strength of the intensity $I$. From Fig.~\ref{fig:main}(e),
we have $g_0^{(2)} \approx 1.97$ and there is an approximately linear
dependence of $P(n_{ph})$ on $n_{ph}$ on the semi-logarithmic scale,
indicating a Bose-Einstein photon number distribution. The dashed-dot
green fitting line for the exponential intensity distribution agrees with
the blue bar. In Fig.~\ref{fig:main}(f), we have $g_0^{(2)}\approx 2.68$,
the dependence of $P(n_{ph})$ on $n_{ph}$ on the semilogarithmic scale
indicates a superthermal distribution, where the dashed magenta fitting
curve obeys the Gaussian-square distribution.

{\bf\it Conclusions}. 
By developing a form of the coupled mode equations for non-Hermitian 
disordered waveguides systems, we have addressed the question of 
whether the recently discovered phenomenon of chiral symmetry protected
thermalization gap can arise in non-Hermitian waveguide systems. We have
found that Hermitian symmetry is a necessary condition for the thermalization 
gap. In particular, we have studied the photon number statistics for two types 
of random disorder: one preserving and another breaking the chiral symmetry,
with the finding that, regardless of whether there is a chiral symmetry, the 
system's being Hermitian is key to the emergence of a photonic thermalization 
gap. Non-Hermitian photonic systems with gain and loss are typical in realistic
situations. For such systems the underlying photon statistics will become
progressively random with the disorder level, eliminating the possibility
of the emergence of any photonic thermalization gap.

Previously, non-Hermitian effects are incorporated into the coupled mode
equations in a phenomenological way~\cite{GWJNLBEMS:2014}, where even
identical homogeneous non-Hermitian waveguides possess a complex coupling
coefficient. We have gone beyond the configuration of identical
homogeneous waveguides and derived a more general form of the non-Hermitian
coupled mode equations. While the focus of our work is on disorder effects,
our general coupled mode equations are suitable for investigating other
system configurations, insofar the coupled mode and small perturbation
approximations hold.

\acknowledgments
We would like to acknowledge support from the Vannevar Bush
Faculty Fellowship program sponsored by the Basic Research Office of
the Assistant Secretary of Defense for Research and Engineering and
funded by the Office of Naval Research through Grant No.~N00014-16-1-2828.


\begin{thebibliography}{10}
\expandafter\ifx\csname url\endcsname\relax\def\url#1{\texttt{#1}}\fi

\bibitem{CW:2015}
\Name{Cao H. \and Wiersig J.} \REVIEW{Rev. Mod. Phys.}{87}{2015}{61}.

\bibitem{KYZ:2016}
\Name{Konotop V.~V., Yang J. \and Zezyulin D.~A.} \REVIEW{Rev. Mod.
  Phys.}{88}{2016}{035002}.

\bibitem{CEPH:2013}
\Name{del Campo A., Egusquiza I.~L., Plenio M.~B. \and Huelga S.~F.}
  \REVIEW{Phys. Rev. Lett.}{110}{2013}{050403}.

\bibitem{RMBNOCP:2013}
\Name{Regensburger A., Miri M.-A., Bersch C., N\"ager J., Onishchukov G.,
  Christodoulides D.~N. \and Peschel U.} \REVIEW{Phys. Rev.
  Lett.}{110}{2013}{223902}.

\bibitem{LRM:2014}
\Name{Lee T.~E., Reiter F. \and Moiseyev N.} \REVIEW{Phys. Rev.
  Lett.}{113}{2014}{250401}.

\bibitem{LC:2014}
\Name{Lee T.~E. \and Chan C.-K.} \REVIEW{Phys. Rev. X}{4}{2014}{041001}.

\bibitem{MPS:2015}
\Name{Malzard S., Poli C. \and Schomerus H.} \REVIEW{Phys. Rev.
  Lett.}{115}{2015}{200402}.

\bibitem{SGWC:2015}
\Name{Sarma R., Ge L., Wiersig J. \and Cao H.} \REVIEW{Phys. Rev.
  Lett.}{114}{2015}{053903}.

\bibitem{L:2016}
\Name{Lee T.~E.} \REVIEW{Phys. Rev. Lett.}{116}{2016}{133903}.

\bibitem{EHWSDCNS:2013}
\Name{Eichelkraut T., Heilmann R., Weimann S., St{\"u}tzer S., Dreisow F.,
  Christodoulides D., Nolte S. \and Szameit A.} \REVIEW{Nat.
  Commun.}{4}{2013}{}.

\bibitem{GWJNLBEMS:2014}
\Name{Golshani M., Weimann S., Jafari K., Nezhad M.~K., Langari A., Bahrampour
  A., Eichelkraut T., Mahdavi S. \and Szameit A.} \REVIEW{Phys. Rev.
  Lett.}{113}{2014}{123903}.

\bibitem{GEBLFBKSHY:2015}
\Name{Gao T., Estrecho E., Bliokh K., Liew T., Fraser M., Brodbeck S., Kamp M.,
  Schneider C., H{\"o}fling S., Yamamoto Y. \etal}
  \REVIEW{Nature}{526}{2015}{554}.

\bibitem{LGCSTR:2012}
\Name{Liertzer M., Ge L., Cerjan A., Stone A.~D., T\"ureci H.~E. \and Rotter
  S.} \REVIEW{Phys. Rev. Lett.}{108}{2012}{173901}.

\bibitem{AHN:2016}
\Name{Amir A., Hatano N. \and Nelson D.~R.} \REVIEW{Phys. Rev.
  E}{93}{2016}{042310}.

\bibitem{KAS:2015}
\Name{Kondakci H.~E., Abouraddy A.~F. \and Saleh B.~E.} \REVIEW{Nat.
  Phys.}{11}{2015}{930}.

\bibitem{KAS2:2015}
\Name{Kondakci H.~E., Abouraddy A.~F. \and Saleh B.~E.}
  \REVIEW{Optica}{2}{2015}{201}.

\bibitem{KSACS:2016}
\Name{Kondakci H.~E., Szameit A., Abouraddy A.~F., Christodoulides D.~N. \and
  Saleh B. E.~A.} \REVIEW{Optica}{3}{2016}{477}.

\bibitem{KKSACS:2016}
\Name{Kondakei A.~E., Keil R., Szameit A., Abouraddy A.~F., Christodoulides
  D.~N. \and Saleh B.~A.} \Book{Tailoring photon-number distribution in
  disordered lattices with chiral symmetry} presented at \Book{2016 CLEO
  (Lasers and Electro-Optics) Conference} (IEEE) 2016.

\bibitem{KAS:2017}
\Name{Kondakci H.~E., Abouraddy A.~F. \and Saleh B. E.~A.} \REVIEW{Sci.
  Rep.}{7}{2017}{8948}.

\bibitem{Szameit:2015}
\Name{Szameit A.} \REVIEW{Nat. Phys.}{11}{2015}{895}.

\bibitem{MECM:2008}
\Name{Makris K.~G., El-Ganainy R., Christodoulides D.~N. \and Musslimani Z.~H.}
  \REVIEW{Phys. Rev. Lett.}{100}{2008}{103904}.

\bibitem{MAKKC:2012}
\Name{Miri M.-A., Aceves A.~B., Kottos T., Kovanis V. \and Christodoulides
  D.~N.} \REVIEW{Phys. Rev. A}{86}{2012}{033801}.

\bibitem{LHZQXKL:2013}
\Name{Luo X., Huang J., Zhong H., Qin X., Xie Q., Kivshar Y.~S. \and Lee C.}
  \REVIEW{Phys. Rev. Lett.}{110}{2013}{243902}.

\bibitem{KMKPSACS:2016}
\Name{Kondakci H.~E., Martin L., Keil R., Perez-Leija A., Szameit A., Abouraddy
  A.~F., Christodoulides D.~N. \and Saleh B. E.~A.} \REVIEW{Phys. Rev.
  A}{94}{2016}{021804}.

\end{thebibliography}

\end{document}